\documentclass[
    ,final            
  ]
  {aipproc}

\layoutstyle{6x9}

\usepackage{bm}

\begin{document}

\title{Chiral Mass Splitting for $\bm{c \bar{s}}$ and $\bm{c \bar{n}}$
 Mesons in the $\bm{\widetilde{U}(12)}$-Classification Scheme of Hadrons}

\classification{}
\keywords{}

\author{Kenji Yamada}{
  address={Department of Engineering Science, Junior College Funabashi
  Campus, Nihon University, Funabashi 274-8501, Japan}
}

\author{Tomohito Maeda}{
  address={Department of Engineering Science, Junior College Funabashi
  Campus, Nihon University, Funabashi 274-8501, Japan}
}

\author{Shin Ishida}{
  address={Research Institute of Science and Technology,
  College of Science and Technology, Nihon University,
  Tokyo 101-8308, Japan}
}

\begin{abstract}
We investigate the chiral mass splitting of parity-doubled $J=0,1$ states
 for $c \bar{s}$ and $c \bar{n}$ meson systems in the
 $\widetilde{U}(12)_{SF}$-classification scheme of hadrons,
 using the linear sigma model to describe the light-quark pseudoscalar
 and scalar mesons
 together with the spontaneous breaking of chiral symmetry, and
 consequently predict the masses of as-yet-unobserved $(0^{+},1^{+})$
 $c \bar n$ mesons. We also mention some indications of their existence
 in the recent published data from the Belle and BABAR Collaborations.
\end{abstract}

\maketitle


\section{Introduction}

Recently, Ishida et al. have proposed the covariant
 $\widetilde{U}(12)_{SF}$-classification scheme of hadrons\cite{IIM2000},
 which gives covariant quark representations for composite hadrons with
 definite Lorentz and chiral transformation properties.
 The $\widetilde{U}(12)_{SF}$-classification scheme has a unitary symmetry
 in the hadron rest frame, called ``static $U(12)_{SF}$
 symmetry''\cite{IIYMO2005},
 embedded in the covariant $\widetilde{U}(12)_{SF}$-representation space,
 of which tensors can be decomposed into representations of
 $\widetilde{U}(4)_{DS} \times SU(3)_{F}$,
 $\widetilde{U}(4)_{DS}$ being the
 pseudounitary homogeneous Lorentz group for Dirac spinors.
 The static $U(12)_{SF}$ contains the Dirac spin group $U(4)_{DS}$
 in its subgroups and
 $U(4)_{DS}$ contains two $SU(2)$ subgroups as
 $U(4)_{DS} \supset SU(2)_{\rho} \times SU(2)_{\sigma}$,
 where $SU(2)_{\rho}$ and $SU(2)_{\sigma}$ are the spin groups
 concerning the boosting and intrinsic-spin rotation, respectively,
 of constituent quarks, being connected with decomposition of Dirac
 $\gamma$-matrices, $\gamma \equiv \rho \otimes \sigma$.
 Thus the static $U(12)_{SF}$ symmetry includes
 the chiral $SU(3)_{L} \times SU(3)_{R}$ symmetry as
 $U(12)_{SF} \supset SU(3)_{L} \times SU(3)_{R} \times SU(2)_{\sigma}$.
 This implies
 that the $\widetilde{U}(12)_{SF}$-classification scheme is able to
 incorporate effectively the effects of chiral symmetry and its
 spontaneous breaking, essential for understanding of properties
 of the low-lying hadrons, into what is called a constituent quark
 model.

\section{Experimental Candidates for the Ground-State
 Quark-Antiquark Mesons}

An essential feature of the $\widetilde{U}(12)_{SF}$-classification
 scheme is to have the static $U(4)_{DS}$ symmetry
 for light $u, d, s$ quarks confined inside hadrons.
 The degree of freedom on the $\rho$-spin, being indispensable
 for covariant description of spin $1/2$ particles, offers a basis
 to define the rule of chiral transformation for quark-composite hadrons.
 Since we have the $\rho$-spin degree of freedom, which is discriminated
 by the eigenvalues of $\rho _{3}$, $r=\pm$, in addition to the ordinary
 $\sigma$-spin, the ground states of light-quark $q \bar{q}$ mesons are
 composed of eight $SU(3)_{F}$ multiplets with respective $J^{PC}$
 quantum numbers, two pseudoscalars
 $(0^{-+}_{\mathrm{N}}, 0^{-+}_{\mathrm{E}})$, two scalars
 $(0^{++}_{\mathrm{N}}, 0^{+-}_{\mathrm{E}})$, two vectors
 $(1^{--}_{\mathrm{N}}, 1^{--}_{\mathrm{E}})$, and two axial-vectors
 $(1^{++}_{\mathrm{N}}, 1^{+-}_{\mathrm{E}})$
 (N and E denoting ``normal'' and ``extra''),
 where each N (E) even-parity multiplet is the chiral partner
 of the corresponding N (E) odd-parity multiplet and
 they form linear representations of the chiral symmetry.
 
Since the eigenstates only with the $\rho _{3}$-eigenvalue
 $r=+$ are taken for heavy quarks, we have
 for heavy-light meson systems two heavy-spin multiplets,
 $(0^{-}, 1^{-})$ and $(0^{+}, 1^{+})$, which are the chiral partner
 each other, while for heavy-heavy meson systems
 we have the same $(0^{-}, 1^{-})$-spin multiplets as in the
 conventional nonrelativistic quark model.

\subsection{The $\widetilde{U}(12)_{SF}$-scheme assignments
 for the observed mesons}

We try to assign some of the observed mesons to the predicted
 $q \bar{q}$ multiplets, resorting to their $J^{PC}$ quantum
 numbers and masses. The observed meson data are taken from
 the Particle Data Group 2004 edition\cite{PDG2004},
 except for the following mesons:

\begin{itemize}

\item $\rho(1250)$.
 There are several experimental indications of the existence
 of the $\rho(1250)$
 reported by the OBELIX\cite{OBELIX1997} and LASS\cite{LASS1994}
 Collaborations, and others.\footnote{See the $\rho(1450)$
 Particle Listings and the ``Note on the $\rho(1700)$''
 in \cite{PDG2004}.}
 
\item $\omega(1200)$.
 The existence of $\omega(1200)$ is claimed in the analysis
 of the $e^{+}e^{-} \rightarrow \pi^{+}\pi^{-}\pi^{0}$
 cross section by the SND Collaboration\cite{SND1999}.
 
\end{itemize}
We accept the existence of these vector mesons as true\cite{Komada}.
 The resulting assignments, though some of them are ambiguous,
 are shown in Table 1.
\begin{table}
    \caption{Experimental candidates for ground-state mesons in the
             $\widetilde{U}(12)_{SF}$-classification scheme.}
	\includegraphics[width=.99\textwidth]{table-1.epsi}
\end{table}
Here we make some comments on these assignments.
\begin{enumerate}
\renewcommand{\labelenumi}{(\theenumi)}
\item The light scalar mesons
 $\{a_{0}(980), \sigma, f_{0}(980), \kappa\}$ are assined to
 the $(0^{++}_{\mathrm{N}})$-nonet as a chiral partner of
 the $\pi$-meson $(0^{-+}_{\mathrm{N}})$-nonet.

\item The low-mass vector mesons
 $\{\rho(1250), \omega(1200), K^{\ast}(1410)\}$ are assined
 to the $(1^{--}_{\mathrm{E}})$-nonet as a chiral partner
 of the $(1^{+-}_{\mathrm{E}})$-nonet
 $\{b_{1}(1235), h_{1}(1170),$ $h_{1}(1380),$ $K_{1}(1400)\}$.

\item The axial-vector mesons
 $\{a_{1}(1260), f_{1}(1285), f_{1}(1420), K_{1}(1270)\}$
 are assined to the $(1^{++}_{\mathrm{N}})$-nonet
 as a chiral partner of the $\rho(770)$-meson
 $(1^{--}_{\mathrm{N}})$-nonet.

\item The recent observed mesons
 $\{D_{sJ}^{\ast}(2317), D_{sJ}(2460)\}$ are assined to
 the $(0^{+}, 1^{+})$ multiplet as a chiral partner of
 the $(0^{-}, 1^{-})$ multiplet $\{D_{s}, D_{s}^{*}\}$\cite{Ishida2003}.
 These newly observed mesons, together with the $\sigma$-meson
 nonet, are the best candidates for the hadronic states with $r=-$
 whose existence is expected
 in the $\widetilde{U}(12)_{SF}$ scheme.

\item It is noted that the normal (N) and extra (E) states
 with the same $J^{PC}$ generally mix together due to
 the spontaneous as well as explicit breaking of chiral symmetry
 and some other mechanism.

\end{enumerate}

\section{CHIRAL MASS SPLITTING FOR THE CHARMED AND
 CHARMED-STRANGE MESON SYSTEMS}

In the $\widetilde{U}(12)_{SF}$-classification scheme
 heavy-light $(c \bar{q})$ meson fields, aside from the
 internal space-time wave functions, are given by
\begin{equation}
	\Phi(v)=\frac{1}{2\sqrt{2}}(1-i v \cdot \gamma)
		(i \gamma _{5} \mathbf{D}
		+i \tilde{\gamma} _{\mu} \mathbf{D}_{\mu}^{\ast}
		+\mathbf{D}_{0}
		+i \gamma _{5}\tilde{\gamma}_{\mu} \mathbf{D}_{1 \mu})
\end{equation}
with
	$v_{\mu} \equiv P_{\mu}/M,\ 
	\tilde{\gamma} _{\mu} \equiv \gamma _{\mu}
		+ v_{\mu}(v \cdot \gamma)$,
where $(\mathbf{D},\mathbf{D}_{\mu}^{\ast},\mathbf{D}_{0},
\mathbf{D}_{1 \mu})$ represent the local fields for the
 $c \bar{q}$ mesons with $J^{P}=(0^{-},1^{-},0^{+},1^{+})$,
 $P_{\mu}$ ($M$) is the four-momentum (mass) of meson fields,
 and flavor indices are omitted for simplicity.
To describe the light-quark pseudoscalar and scalar mesons
 together with the spontaneous breaking of chiral symmetry,
 we adopt the $SU(3)$ linear sigma model, introducing
 the chiral field $\Sigma _{5}$ defined by
\begin{equation}
	\Sigma _{5} = s-i \gamma _{5}\phi
\end{equation}
with
	\[ s=\frac{1}{\sqrt{2}} s^{a} \lambda^{a}, \ \
	\phi=\frac{1}{\sqrt{2}} \phi^{a} \lambda^{a} \ \ \
	(a=0, \cdots ,8), \]
where $\lambda^{0}=\sqrt{2/3} \ \mathbf{1}$ and $s^{a}$
 ($\phi^{a}$)
 are the scalar (pseudoscalar) fields. We now write
 a chiral-symmetric effective Lagrangian which gives
 the chiral mass splitting between the heavy-light
 $(0^{-},1^{-})$ and $(0^{+},1^{+})$ multiplets
 through the spontaneous breaking of
 chiral symmetry\cite{BEH2003,Ishida2003}:
\begin{equation}
	\mathcal{L}_{ND}=-g_{ND} \mathrm{Tr}
	[\Phi \Sigma _{5} \bar{\Phi}] \label{eq:LND},
\end{equation}
where $g_{ND}$ is the dimensionless coupling constant
 of Yukawa interaction in the nonderivative form and
 the trace is taken over the spinor and flavor indices.
 
When the chiral symmetry is spontaneously broken,
 $s$ has the vacuum expectation value,
 $\langle s \rangle _{0}=\mathrm{diag}(a,a,b)$,
 where $a$ and $b$ are related to the pion and kaon decay
 constants by
\begin{equation}
	a=\frac{1}{\sqrt{2}} f_{\pi}, \ \
	b=\frac{1}{\sqrt{2}} (2f_{k}-f_{\pi}).
\end{equation}
Then the mass splitting between the two multiplets
 is induced and the mass differences $\Delta M_{\chi}(c \bar q)$
 are given by
	$\Delta M_{\chi}(c \bar n)=2g_{ND} a$ and 
	$\Delta M_{\chi}(c \bar s)=2g_{ND} b$,
which leads to the relation
\begin{equation}
	\Delta M_{\chi}(c \bar n)=\Delta M_{\chi}(c \bar s) \frac{a}{b}
	=\Delta M_{\chi}(c \bar s) \left( \frac{2f_{K}}{f_{\pi}}-1 \right)^{-1}.
\end{equation}
From this relation with the experimental values\cite{PDG2004},
	$\Delta M_{\chi}(c \bar s) = 348.0 \pm 0.8 \ \mathrm{MeV}$
	and
	$f_{K^{+}}/f_{\pi^{+}} = 1.223 \pm 0.015$,
we obtain
	$\Delta M_{\chi}(c \bar n) = 240.8 \pm 5.4 \ \mathrm{MeV}$
and consequently predict the masses
\begin{equation}
	M(D_{0}^{\ast}) = 2.11 \pm 0.01 \ \mathrm{GeV}, \ \
	M(D_{1}) = 2.25 \pm 0.01 \ \mathrm{GeV}
\end{equation}
for the $(0^{+},1^{+})$ $c \bar n$ mesons, using the measured
 mass values\cite{PDG2004} of the $D(0^{-})$ and $D^{\ast}(1^{-})$ mesons.
 We hereafter refer to these predicted mesons, respectively, as
 ``$D_{0}^{\ast}(2110)$'' and ``$D_{1}(2250)$''.

\section{POSSIBLE INDICATIONS OF THE EXISTENCE OF LIGHT SCALAR
 AND AXIAL-VECTOR CHARMED MESONS}

We could ask experimental data whether there was some evidence
 for the existence of $D_{0}^{\ast}(2110)$ and $D_{1}(2250)$.
 Here we check on the recent published data on the
 $D \pi$ and $D^{*} \pi$ mass distributions in
 $B \rightarrow (D \pi)\pi$, $(D^{*}\pi)\pi$ decays from
 the Belle\cite{Belle2004} and BABAR\cite{BABAR2003} Collaborations.

\begin{itemize}
\item \textbf{$D \pi$ mass spectrum}:
 In the Belle data\footnote{See the $D \pi$ mass distribution
 in Figure 3 of \cite{Belle2004}.}
 we see an excess of events, a single data point of 20 MeV bin,
 at a mass of 2.13 GeV near the predicted mass of the
 $D_{0}^{\ast}(2110)$, and so might regard it as an indication
 of that resonance, though it is natural to think that
 its data point should be within a statistical error.
 On the other hand, it would seem to us that the BABAR
 data\footnote{See the $D \pi$ mass distribution
 in Figure 3 (right) of \cite{BABAR2003}.} around a mass of
 2.1 GeV show a typical pattern of interference between
 two or more resonances.

\item \textbf{$D^{*}\pi$ mass spectrum}:
 In the Belle data\footnote{See the $D^{*} \pi$ mass distribution
 in Figure 9 of \cite{Belle2004}.} there is also an excess of events,
 a single data point of 10 MeV bin, at a mass of 2.255 GeV
 near the predicted mass of the $D_{1}(2250)$,
 and so it might be an indication of the resonance.
 Although it is not clear, the BABAR data\footnote{See the $D^{*} \pi$
 mass distribution in Figure 3 (left) of \cite{BABAR2003}.}
 around a mass of 2.26 GeV might show a typical pattern
 of interference.

\end{itemize}

If the $D_{0}^{\ast}(2110)$ and $D_{1}(2250)$ resonances
 really exist, their widths have to be narrow,
 $\leq 20 \textrm{-} 30$ MeV, judging from the data mentioned above.
 The dominant decay modes of these resonances are
 $D \pi$ and $D^{*} \pi$, respectively, and thus
 we examine their single pion transitions.
 To estimate the widths of $D_{0}^{\ast}
 \rightarrow D + \pi$ and $D_{1}
 \rightarrow D^{*} + \pi$ decays, together with
 $D^{\ast} \rightarrow D + \pi$, we set up,
 in addition to the nonderivative interaction $\mathcal{L}_{ND}$
 in Eq. (\ref{eq:LND}),
 the chiral-invariant effective interaction
 with the derivative form:
\begin{equation}
	\mathcal{L}_{D} = g_{D} \mathrm{Tr}
	[\Phi (\partial _{\mu} \Sigma _{5}) \gamma _{\mu}
	(F_{U} \bar{\Phi})] \label{eq:LD},
\end{equation}
where
 $F_{U} = \gamma \cdot \partial
 / \sqrt{\partial \cdot \partial}$
 and $g_{D}$ is the coupling constant with a dimension
 of $(mass)^{-1}$, which is related to the axial coupling
 constant $g_{A}$ by
 $g_{D} = g_{A}/2a = g_{A}/\sqrt{2} f_{\pi}$.
 The pionic decay widths of the $D^{*}$, $D_{0}^{\ast}$,
 and $D_{1}$ states are derived from $\mathcal{L}_{ND}$
 and $\mathcal{L}_{D}$, and the decay widths of
 $D_{0}^{\ast}$ and $D_{1}$ are identical. Using the measured
 value of $\Gamma [D^{*+} \rightarrow D^{0} + \pi ^{+}]
 =65 \ \mathrm{keV}$\cite{PDG2004}
 and $g_{ND}=1.84$ from $\Delta M_{\chi}(c \bar n)
 = 241 \ \mathrm{MeV}$,
 the coupling $g_{D}$ is fixed to 3.96 GeV$^{-1}$
 (corresponding to $g_{A}=0.521$), and then we obtain
\begin{equation}
	\Gamma [D_{0}^{*}(2110) \rightarrow D + \pi]
	=\Gamma [D_{1}(2250) \rightarrow D^{*} + \pi]
	\approx 30 \ \mathrm{MeV}.
\end{equation}
This value is consistent with the speculated widths of
 the $D_{0}^{\ast}(2110)$ and $D_{1}(2250)$.

\section{CONCLUDING REMARKS}

We have presented the possible assignments for some of
 the observed mesons in the covariant
 $\widetilde{U}(12)_{SF}$-classification scheme.
 It is necessary and important to examine the strong- and
 radiative-decay\cite{Maeda} properties of the assigned states
 in order to establish their assignments.
 On the basis of these assignments we have also predicted
 the existence of the low-mass
 $(0^{+},1^{+})$ $c \bar n$ mesons with narrow width,
 which might have been seen in the recent published data
 on the $D \pi$ and $D^{*} \pi$ mass distributions
 from the Belle and BABAR Collaborations.

\bibliographystyle{aipprocl} 

\begin{thebibliography}{20}

\bibitem{IIM2000}
S. Ishida, M. Ishida, and T. Maeda,
 \emph{Prog. Theor. Phys.} \textbf{104}, 785 (2000).

\bibitem{IIYMO2005}
S. Ishida, M. Ishida, K. Yamada, T. Maeda, and M. Oda,
 hep-ph/0408136.

\bibitem{PDG2004}
Particle Data Group, S. Eidelman et al.,
 \emph{Phys. Lett.} \textbf{B592}, 1 (2004).

\bibitem{OBELIX1997}
OBELIX Collaboration, A. Bertin et al.,
 \emph{Phys. Lett.} \textbf{B414}, 220 (1997).
 
\bibitem{LASS1994}
LASS Collaboration, D. Aston et al.,
 SLAC-PUB-5606 (1994).

\bibitem{SND1999}
M. N. Achasov et al.,
 \emph{Phys. Lett.} \textbf{B462}, 365 (1999).

\bibitem{Komada}
T. Komada, these proceedings.

\bibitem{Ishida2003}
S. Ishida,
 in \emph{HADRON SPECTROSCOPY}, edited by E. Klempt et al.,
 AIP Conference Proceedings 717, Melville, New York,
 2004, p. 716.

\bibitem{Belle2004}
Belle Collaboration, K. Abe et al.,
 \emph{Phys. Rev. D} \textbf{69}, 112002 (2004).

\bibitem{BABAR2003}
BABAR Collaboration, B. Aubert et al.,
 BABAR-CONF-03/020, SLAC-PUB-10107 (2003), hep-ex/0308026.

\bibitem{BEH2003}
W. A. Bardeen, E. J. Eichten, and C. T. Hill,
 \emph{Phys. Rev. D} \textbf{68}, 054024 (2003).\\
 M. A. Nowak, M. Rho, and I. Zahed,
 \emph{Acta Phys. Polon.} \textbf{B35}, 2377 (2004).

\bibitem{Maeda}
T. Maeda, K. Yamada, M. Oda, and S. Ishida,
 these proceedings.

\end{thebibliography}

\end{document}